\magnification=\magstep1
\tolerance=500
\vskip 2 true cm
\rightline{TAUP 2760-04}
\rightline{15 March, 2004}
\bigskip
\centerline{\bf Stark Effect in Lax-Phillips Scattering Theory}
\bigskip
\centerline{Tamar Ben Ari$^1$ L.P. Horwitz$^{1,2}$}
\smallskip
\centerline{$^1$ Department of Physics, Bar Ilan University}
\centerline{Ramat Gan 52900, Israel} 
\centerline{$^2$ School of Physics, Tel Aviv University}
\centerline{Ramat Aviv 69978, Israel}
\bigskip
\noindent {\it Abstract\/}: The scattering theory of Lax and Phillips,
 originally developed to describe resonances associated with
 classical wave equations, has been recently extended to apply as well
 to the case of the Schr\"odinger equation in the case that the wave
 operators for the corresponding Lax-Phillips theory exist. It is
 known that the bound state levels of an atom become resonances
 (spectral enhancements) in the continuum in the presence of an
 electric field (on all space) in the quantum mechanical Hilbert
 space.  Such resonances appear as states in the extended Lax-Phillips
 Hilbert space.  We show that for a simple version of the Stark
 effect, these states can be explicitly computed, and exhibit the
 (necessarily) semigroup property of decay in time.  The widths and
 location of the resonances are those given by the poles of the
 resolvent of the standard quantum mechanical form.
\vfill  
\break

\bigskip
\noindent
{\bf 1. Introduction}
\par The description commonly used for an unstable system is that of
Wigner and Weisskopf$^1$. They assumed that the unstable system is
 represented by a vector, say $\phi$, in the Hilbert space of states,
 and that the Hamiltonian evolution of this state permits the
 development of other components which represent the ``decayed''
 system.  One may assume a Hamiltonian $H_0$ for which $\phi$ is an
 eigenstate, and the full Hamiltonian is constructed by adding a
 pertubation $V$ which induces transitions from this state. The
 basic assumption of their physical model is that these transitions
 carry the state of the system from that of some specific system to a
 state in the continuous spectrum which contains the decay products.
 For example, one may think of an atom in an excited state as the
 unstable system; in the absence of electromagnetic interaction, this
 state would be a stable bound state.  A term added to the Hamiltonian
 corresponding to electromagnetic interaction induces a transition
 from this state, and the square of the corresponding transition
 amplitudes,
 which may be generally computed perturbatively, give the probability for
 decay.  This very standard technique is rigorously correct for
 reversible quantum transitions, according to the laws of quantum
 theory. When applied to the decay of an unstable system, however, for
 which the evolution is irreversible, it consitutes an
 approximation which may be inadequate.  In the following we discuss
 some of the difficulties of the application of the Wigner-Weisskopf
 method to the treatment of irreversible phenomena, such as particle
 decay, and in Section 3 we describe the general structure of the
 Lax-Phillips theory, which yields an exact semigroup evolution and
 appears to be a more appropriate description of such phenomena.  In
 Section 2, we apply the Wigner-Weisskopf theory to a simple model for
 the Stark effect, for comparison, and work out the Lax-Phillips
 theory for this model in Section 4.  Some of the calculations in
 Section 2 coincide with those needed in Section 4; the approximate
 pole approximation decay law of the Wigner-Weisskopf theory contains
 exactly the same pole as the singularity of the Lax-Phillips $S$
 matrix, so that the decay laws are (approximately) the same, but, as
 we shall see, there are important differences in that the
 Lax-Phillips treatment leads to an exact semigroup and contains
 information not accessible in the Wigner-Weisskopf method.   
\par One easily sees that the Wigner-Weisskopf procedure cannot yield
 a semigroup property
 of evolution. The ``survival amplitude'' for such a state is given by
$$        A(t) = (\phi, e^{-iHt}\phi), \eqno(1.1)$$
  and the survival probability is $|A(t)|^2$.  One observes, in
  general, that decaying systems are characterized by exponential
  decay in time. However, the time derivative of $|A(t)|^2$ at $t=0$
  is
$$ \eqalign{{d \over dt}|A(t)|^2|_{t=0} &= -i(\phi,H\phi)
  +i(H\phi,\phi)\cr &= 0, \cr} \eqno(1.2)$$
where the last equality follows from the self-adjointness of $H$.
Hence the probability curve must arrive at the origin with zero slope,
and cannot be exponential.  Generally, this deviation from exponential
occurs only for very short times $^{2,3}$, e.g., for a weak decay with
lifetime of the order of $10^{-10} - 10^{-8}$ seconds, this short time
may be of order $10^{-22}$ seconds.  The dominating exponential
behavior for intermediate times can well account for the data in many
such cases of single channel decay.  For multichannel problems (many
types of final states), for example, the two channel neutral K-meson
decay, the lack of semigroup property may lead to disagreement with
experiment.
\par To see this (and to see as well the origin of the exponential
part of the decay), let us examine briefly the two channel case. In
this case, the probability of decay is given by
$$ P =  \sum_j \int{ d\lambda_j \vert <\lambda_j \vert e^{-iHt}\phi) \vert^2},
\eqno(1.3)$$
where $\{\lambda_j\}$ corresponds to the continuous spectrum of the
``final'' states, and $\phi$ is an arbitrary state in the two
dimensional subspace of initial states (e.g., $K^0, {\overline K}^0$).
This can be written alternatively, since the evolution is unitary, as 
$$P = 1 -  \sum_j \vert(\phi_j, e^{-iHt}\phi)\vert^2, $$
where $\{\phi_j\}$ spans the subspace.
Since one can always express $\phi$ as a linear superposition of the
$\{\phi_j\}$, it is sufficient to study the quantities
$$ \eqalign{ A_{ij} &= (\phi_i, e^{-iHt} \phi_j)\cr
 &= {1 \over 2\pi i}\int_C e^{-izt}(\phi_i, {1 \over z-H} \phi_j) dz\cr
&=  {1 \over 2\pi i}\int_C e^{-izt}\bigl({1 \over z-
W(z)}\bigr)_{ij}dz,\cr}
 \eqno(1.4)$$
where $W(z)$ is a $2 \times 2$ matrix defined by the second
equality, $C$ is a contour for the inverse Laplace transform, and the
quantity $(z-H)^{-1}$ is the resolvent operator associated with the
Hamiltonian $H$ (defined for $z$ in the upper half plane). The contour
$C$ runs above the real line from
$+\infty$ to $-\infty$. Almost every $2 \times 2$ matrix has a
 decomposition$^4$ of spectral type 
$$ W(z) = z_1(z) Q_1(z) + z_2(z)Q_2(z), \eqno(1.5)$$
where the $Q_i(z)$ are constructed from the right and left
eigenvectors of $W(z)$, and satisfy $Q_i^2 = Q_i, \,\, Q_i Q_j =0, i
\neq j$. The resolvent kernel is regular in the upper half plane, but
its analytic continuation through the cut induced by the continuous
spectrum of $H$ may have poles (or, in principle, other
singularities).  For small perturbations, one expects that the bound
states in $H_0$ for a model for which $H = H_0 + V$ would move
slightly into this region, which we shall call the second Riemann 
sheet$^4$.  Poles would appear as
zeros in either $z-z_1(z)$ or in $z-z_2(z)$, at points $z_a$ or $z_b$.
Let us assume that the eigenvalues are ordered so that one pole
appears at $z_a$ such that $z_a -z_1(z_a)=0$ and the second at $z_b$
such that  $z_b -z_1(z_b)=0$. In distorting the contour below the real
axis across the cut into the second sheet, these pole contributions
may dominate the intermediate time behavior of the amplitude (the very
short time behavior is controlled by the Taylor expansion at small
$t$)$^3$.  For semibounded spectrum, in $(0, \infty)$, the long time
behavior is controlled by the contribution at the branch point, and
goes as $t^{-n}$, where $n$ is the dimensionality of space.  The
resulting pole approximation is then
$$ A_{ij}(t) \cong  e^{-iz_a t} Q_1(z_a) + e^{-iz_b t}Q_2 (z_b),
\eqno(1.6)$$
where we have also approximated the factor in the residue arising from the
derivative of the denominator in the neighborhood of the singularity
by unity. 
     \par For the single channel case, one sees the emergence of the
simple exponential behavior for intermediate times.  However, for the
two channel (or more) case, we see that the matrix $A_{ij}(t)$ does
not satisfy the semigroup property
$$ A(t)A(t') = A(t+t') \eqno(1.7)$$
because, in general (in the absence of CP conservation),
$Q_1(z_a)Q_2(z_b) \neq 0$ (in case $W(z)$ is independent of $z$, these 
 matrices would be, of  course, orthogonal).    
\par One may think of this semigroup property as a requirement for the
evolution of an irreversible system$^5$; this follows from arguing
that the evolution on a Hilbert space must be composable (in the sense
that a product of evolutions must be equivalent to a single evolution
over the total interval) for a
sequence of $t$'s that are positive, since the system carries no
dependence on its past, but the inverse does not exist.  Indeed, 
experiments carried out at Fermilab$^6$ show that the phenomenological
model of Lee, Oehme, Yang and Wu$^7$ is an extremely accurate
representation.  This model assumes that the evolution in the two
dimensional subspace is given by $\exp(-iH_{2 \times 2}t)$, where
$H_{2 \times 2}$    is a $2 \times 2$ non-Hermitian matrix. In this
case, the matrix $H_{2 \times 2}$ , independent of $z$, replaces $W(z)$ in the 
 the spectral decomposition $(1.5)$; the pole residues are orthogonal,
and one obtains the exact semigroup law.  Using a Lee-Friedrichs
model$^4$ to estimate the errors, one finds that the non-orthogonality
of $Q(z_a)$ and $Q(z_b)$ induces errors that are larger than the 
experimental errors in
regeneration experiments and in the interference in the exit
beam$^8$. We infer from these results that for irreversible
transitions (of the type occurring in particle decay), one should have
the semigroup property.   
\par The theory of scattering developed in 1967 by Lax and Phillips$^9$
 for application to classical wave equations has the property
that resonance decay satisfies exactly the semigroup property. It is
also true that, in that theory, there is a well-defined state in the
 Hilbert space corresponding to the resonance. 
\par This theory has been developed and methods found to apply it to
resonances in the quantum theory$^{10-14}$.  In the Wigner-Weisskopf
method, the resonance is represented only as a pole in the complex
plane. Attempts have been made to extend the Hilbert space to the
so-called rigged Hilbert space$^{15}$ (Gel'fand triple constuction), in which
exact exponential decay occurs, but the ``eigenstate''
of the resonance, in that case, lies in a Banach space that is not
generally a Hilbert space, so that scalar products and expectation
values are not defined. The definition of the wave function of the
resonance carrying information on its size, shape and momentum
distribution could be  important in many applications, for example, for the
resonance in a quantum dot or a small metallic particle$^{16}$.
\par It is most straightforward
to apply these methods to problems in which the corresponding
Hamiltonian of the standard theory has spectrum $-\infty$ to
$+\infty$.  One such example is the relativistic quantum theory$^{17}$, where
the unperturbed Hamiltonian has a form proportional to $E^2 -p^2$; the
Lax-Phillips theory for such a problem has been worked out$^{18}$.  For
the non-relativistic case, the spectrum of the Hamiltonian is
generally bounded from below, and further studies are being carried
out in order to be able to apply the theory in these cases as well$^{14}$.
However, the non-relativistic Stark effect is a case in which the
non-relativistic problem has a Hamiltonian unbounded from below.  We
shall study this problem in detail in this paper, and show that one
finds semigroup behavior for the decay of the Stark induced
resonances; the wave functions for the resonant states are also explicity 
worked out. In the
following we study a simple model, of the type of Friedrichs and
Rejto$^{19}$, for the Stark effect, in the framework of Wigner and Weisskopf,
and then imbed the results into the more general framework of Lax and
Phillips. We shall review the content of the Lax Phillips theory in
this later section.
\vfill
\break
\bigskip
\noindent
{\bf 2. Wigner-Weisskopf Analysis  of a Stark Model}
\smallskip
\par
The potential for the Stark effect problem, of the form $-Ex$, is
 unbounded (on the full space; we shall work in one space dimension).
   For a model of the form$^{19}$
$$ H = -Ex + \lambda P_0, \eqno(2.1)$$
where $\lambda$ is real and $P_0$ is a rank one projection operator, 
it will be convenient to consider $-Ex \equiv H_0$ as the unperturbed
 Hamiltonian
and  the second term, $\lambda P_0 \equiv V$ as the perturbation (the resolvent
is, of course unaffected by this choice, but the form of the
perturbation theory is very different). We study in
this section the Wigner-Weisskopf description of the resonance, and in
the next section, imbed this analysis in the Lax-Phillips Hilbert
space.
 \par Let us choose for $P_0$ the form
$$ <x | P_0 | x'> =\bigl({2 \over \pi}\bigr)^{1 \over 2}  e^{-(x^2 + x'^2)} \eqno(2.2)$$   
The resolvent satisfies the identity (second resolvent equation)
$$ G= G_0 + G_0 V G, \eqno(2.3)$$
where, as above, $G=(z-H)^{-1}$, and $G_0 = (z-H_0)^{-1}$, defined for
$z$ in the upper half plane,
 where $H_0= -Ex$. The $x,x'$ matrix element of $G$ is therefore
$$ \langle x \vert G \vert x' \rangle = {1 \over z+Ex} \delta(x-x')
+ {1 \over z+Ex}\lambda \int_{-\infty}^{\infty} \langle x \vert P_0
\vert x'' \rangle \langle x'' \vert G \vert x' \rangle dx''. \eqno(2.4)$$
 Let us define
$$ f(z,x') = \int dx'' e^{-x''^2} \langle x'' \vert G \vert x' \rangle
dx''. \eqno(2.5)$$
It then follows from $(2.4)$ that
$$ f(z, x') = \int dx {e^{-x^2} \over z+ Ex} + \lambda \int dx
{e^{-x^2} \over z+ Ex} \sqrt{2 \over \pi} e^{-(x^2 + x''^2)} \langle
x'' \vert G \vert x' \rangle dx''.$$
One can write this as 
$$ f(z,x') =  {e^{-x'^2} \over z+ Ex'}  + \lambda  \sqrt{2 \over \pi}
F(z) f(z,x'), $$
or
$$ f(z,x') = { 1 \over 1- \lambda \sqrt{2 \over \pi} F(z)}
 {e^{-x'^2} \over z+ Ex'}, \eqno(2.6)$$
where 
$$ F(z) = \int\, dx {e^{-2x^2} \over z+ Ex}= {i\pi\over
E}e^{-{2z^2\over E^2}}erfc[i\sqrt{2}{z\over E}]. \eqno(2.7)$$

Returning to Eq. $(2.4)$, we see that
$$ \eqalign{ \langle x \vert G(z) \vert x' \rangle &= {1 \over z+Ex}
\delta(x-x') + {1 \over z+Ex}\lambda \sqrt{ 2 \over \pi} e^{-x^2}
f(z,x') \cr
 &= {1 \over z+Ex}\delta(x-x') + \lambda \sqrt{2 \over \pi} {e^{-x^2}
e^{-x'^2} \over (z+Ex)(z+Ex')} { 1 \over 1- \lambda \sqrt{2 \over \pi}
F(z)}. \cr} \eqno(2.8)$$
\par We now wish to approximate the time behavior of the survival amplitude.
As in Eq.$(1.4)$, the time dependence of the survival amplitude (for
one channel) is given by
$$ A(t) = {1 \over 2\pi i} \int_C (\varphi \vert G(z) \vert \varphi )e^{-izt}
.\eqno(2.9) $$
The contour $C$ corresponds to a line running in the complex energy
plane from right to left slightly above the real axis. The matrix element 
$(\varphi \vert G(z) \vert \varphi )$ is analytic in the upper
half-plane. We can shift this line continuously and differentiably
through the real axis into the lower half plane, provided that the
contribution of the vertical pieces at $\pm \infty$
vanish. It is clear from $(2.7)$ and $(2.8)$ that this is true for the
part of the vertical integrations that lie in the upper half plane. To
write the integrand along the new curve below the axis, we must
analytically continue $F(z)$.  To do this, consider (for $\xi$ real)
$$\eqalign{ F(\xi + i\epsilon) - F(\xi -i\epsilon) &= \int_{-\infty}^\infty
dx e^{-2x^2} \bigl\{ {1 \over \xi + i\epsilon +Ex} - {1 \over \xi -
i\epsilon + Ex}\bigr\} \cr 
&= -2\pi i\int dx e^{-2x^2}\delta(\xi + Ex)  \cr
&= -{2\pi i \over E} e^{-{2\xi^2 \over E^2}} \cr}
 \eqno(2.10)$$
This function has an analytic extension in the finite lower half plane,
given by
 $$ F^\ell (z) = F(z) - {2\pi i \over E} e^{-{2z^2 \over
E^2}}. \eqno(2.11)$$
\par We find, numerically, for a reasonable choice of parameters and a
simple assumption for $\varphi(x)$, that the analytic continuation of
 the function
$$\eqalign{(\varphi \vert G(z) \vert \varphi ) &= \int dx dx'
 \varphi^*(x)\Bigl(
 {1 \over z+Ex}\delta(x-x') \cr &+ \lambda \sqrt{2 \over \pi} {e^{-x^2}
e^{-x'^2} \over (z+Ex)(z+Ex')} { 1 \over 1- \lambda \sqrt{2 \over \pi}
F(z)}\Bigr)\varphi(x),\cr}  \eqno(2.12)$$
 defined by $(2.8)$ and $(2.11)$, into the lower half plane, has a
pole, inducing an exponential decay term to the amplitude. The
function $(\varphi \vert G(z) \vert \varphi )$ has a very simple form
if we assume that $\varphi(x)$ has the Gaussian form
$$ \varphi(x) = \sqrt{2\over \pi} e^{-x^2}. \eqno(2.13)$$
The first term of $(2.12)$ contains $e^{-2x^2}$; its integral
with the denominator $z+Ex$ is, according to $(2.7)$, the function
$F(z)$. The second term factorizes into two integrals of the same
form.  It then follows, with this assumption on $\varphi$, that 
$$\eqalign{(\varphi \vert G(z) \vert \varphi )&= \sqrt{2 \over \pi}
F(z) + \lambda({2 \over \pi}) { (F(z))^2 \over  1- \lambda \sqrt{2 \over \pi}
F(z)} \cr
&= {\sqrt{2 \over \pi} F(z)   \over  1- \lambda \sqrt{2 \over \pi}
F(z)}.\cr} \eqno(2.14)$$
\par The analytic continuation of this function into the lower half
plane is achieved by the continuation of $F(z)$ to $F^\ell(z)$; this
function has no poles in the finite lower half plane, and hence the
pole can only come from the condition
$$  g(z) \equiv 1- \lambda \sqrt{2 \over \pi} F^\ell(z)= 0.\eqno(2.15)$$
\par For the value $\lambda/ E = 11$, MAPLE provides us with a unique
solution for the position of the pole, $z_0 = -4.446 -.31896\times
 10^{-15} i$, which has, as expected, a very small imaginary part for this
 reasonably physical choice of parameters.  Differentiating $(2.15)$
 implicitly with respect to $E$, we find that the real part of the
 pole moves to more negative values as $E$ increases.  Since the
 unpertubed system has mean position of the particle at zero, this
 shift corresponds to an increase in field induced polarization with
 increasing value of the field.
\par  There
remains, however, a contribution to the survival amplitude
from integration on a line running from $+\infty$ to $-\infty$ on the
real part of $z=\xi +i\zeta$, where $\zeta$ can be very large and
negative.  The contribution of this so-called background integral is
strongly suppressed by the exponent $\exp(-izt)$ for t large and
positive. For small $t$, however, this suppression is not strong
unless $\zeta \rightarrow -\infty$.  However, in this limit,
the integrand is not well-defined, since for any large and negative
$\zeta$, the contribution of the discontinuity in $F(z)$ strongly
suppresses the integrand for $\xi$ small compared to $\zeta$.  This
suppression is not maintained, however, for the contributions from
$\xi$ in the neighborhood or greater than $\zeta$.  Hence the
convergence of the contribution on the background is not uniform.  We
see  from the general argument $(1.2)$ that the pole contribution
cannot represent the result precisely for small $t$, and therefore the
Wigner-Weisskopf treatment, even in this case of unbounded spectrum,
cannot result in a pure exponential (semigroup) behavior for the
reduced evolution.  It is exactly in this respect that the
Lax-Phillips treatment provides a result which is closer to the
physics of irreversible decay.
 \bigskip
\noindent
{\bf 3. Lax-Phillips theory}
\smallskip
\par The Stark effect admits a simple transition to the
Lax-Phillips framework, since the Hamiltonian is unbounded.  In the
following, we briefly describe the theory of Lax and Phillips$^{9}$ and its
extension to the quantum case$^{13}$, and then apply the formalism to
the model we are using.
\par The scattering theory of Lax and Phillips assumes the existence
of a Hilbert space $\overline{\cal H}$ of physical states in which
 there are
 two distinguished orthogonal subspaces ${\cal D}_+$ and ${\cal D}_-$ with the
 properties
$$\eqalign{ U(\tau)\,{\cal  D}_+ &\subset \,{\cal D}_+  \qquad \tau > 0 \cr
U(\tau) \,{\cal D}_- &\subset \,{\cal D}_- \qquad \tau < 0\cr
\bigcap_\tau \,U(\tau)\,{\cal D}_\pm &= \{0\} \cr
\overline{\bigcup_\tau \,U(\tau)\,{\cal D}_\pm }
&= \overline {\cal H}, \cr} \eqno(3.1)$$
i.e., the subspaces ${\cal D}_\pm$ are stable under the action of the full
 unitary dynamical evolution $U(\tau)$, a function of the physical
 laboratory time, for positive and negative
times $\tau$ respectively; over all $\tau$, the evolution operator
generates a dense
 set in $\overline{\cal H}$ from either ${\cal D}_+$ or ${\cal D}_-$.
 We shall call ${\cal D}_+$ the {\it outgoing subspace} and
${\cal D}_-$ the {\it incoming subspace} with respect to the group $U(\tau)$.
\par A theorem of Sinai$^{20}$ then assures that $\overline {\cal H}$ can
be represented as a family of Hilbert spaces obtained by foliating
$\overline {\cal H}$ along the real line, which we shall call
$\{s\}$, in the form of a direct integral
$$\overline{\cal H} = \int_\oplus {\cal H}_s,   \eqno(3.2)$$
where the set of auxiliary Hilbert spaces ${\cal H}_s$ are all
isomorphic.  Representing these spaces in terms of square-integrable
functions, we define the norm in the direct integral space (we use
Lesbesgue measure) as
$$ \Vert f \Vert^2 = \int_{-\infty}^\infty \,ds \Vert f_s \Vert^2_H,
\eqno(3.3)$$
where $f \in \overline{H}$ represents
a vector in $\overline {\cal H} $ in terms of  the $L^2$ function space
$L^2(-\infty,\infty, H)$,
and $f_s \in H$, the $L^2$ function space
representing ${\cal H}_s$ for any $s$. The Sinai theorem
 furthermore asserts that
there are representations for which the action of the full
evolution group $U(\tau)$ on $L^2(-\infty,\infty, H)$ is translation
by $\tau$ units. Given $D_\pm$ (the $L^2$ spaces representing ${\cal D}_\pm$),
 there is such a representation, called the
{\it incoming translation representation}$^1$, for which functions in $D_-$ have support
in $L^2(-\infty,0, H)$, and another called the {\it outgoing translation
representation}, for which functions in $D_+$ have support in $L^2(0,
 \infty,H)$.
\par  Lax and Phillips$^9$ show that there are unitary operators
$W_\pm$,
 called
 wave
operators, which map elements in $\overline{\cal H}$,
respectively, to these representations.  They define an $S$-matrix,
$$ S= W_+W_-^{-1} \eqno(3.4)$$
which connects these representations; it is unitary, commutes with
translations, and maps $L^2(-\infty,0)$ into itself\footnote{*}{Note that 
the $S$ matrix of the Lax-Phillips theory maps the incoming to
 the outgoing {\it representations} unlike the $S$ matrix of the usual 
scattering theory which maps incoming to outgoing {\it states}}.  The
singularities of this $S$-matrix, in what we shall define as the {\it
spectral representation}, correspond to the spectrum of the generator
of the exact semigroup characterizing the evolution of the unstable
system.
\par With the assumptions stated above on the properties of the
subspaces ${\cal D}_+$ and ${\cal D}_-$, Lax and Phillips$^9$ prove that the
 family
of operators
$$ Z(\tau) \equiv P_+ U(\tau) P_- \qquad (\tau \geq 0), \eqno(3.5)$$
where $P_\pm$ are projections into the orthogonal complements of
${\cal D}_\pm$, respectively, is a contractive, continuous, semigroup 
satisfying
$$ Z(\tau_1)Z(\tau_2) = Z(\tau_1 + \tau_2) \eqno(3.6)$$
  This operator
 annihilates vectors in ${\cal D}_\pm$ and carries the space
$$ {\cal K} = \overline {\cal H} \ominus {\cal D}_+ \ominus
 {\cal D}_- \eqno(3.7)$$
into itself, with norm tending to zero for every element in
${\cal K}$. One may demonstrate the semigroup property by noting that
the operator acting on a vector in
$D_-$ vanishes by definition; on a vector in $D_+$, it vanishes due to the
 invariance of $D+$. The scalar product on the left with a vector $g$
in $D_+$ vanishes by definiton; if $g$ is in $D_-$, the factor $P_+$ may be
 suppressed, since such a $g$ is in a subspace othogonal to $D_+$.  Hence the 
operator $(3.5)$ can only connect ${\cal K}$ to ${\cal K}$.  In the product
$$ P_+U(\tau_1)P_- P_+U(\tau_2)P_- \eqno(3.8)$$
one may suppress the factor $P_-$ following $P_+$ between the two unitaries 
since vectors in ${\cal K}$ are orthogonal to $D_-$.  Then, consider
$$ P_+ U(\tau_1)P_+ U(\tau_2) = P_+ U(\tau_1)[1 - (1-P_+)]U(\tau_2);$$
the second term in square brackets makes no contribution since it is a
 projection into $D_+$, and $U(\tau_1)$ leaves this subspace
invariant. The composition of $U(\tau_1)$ and $U(\tau_2)$ then proves
$(3.6)$.
  \par  The outgoing subspace $D_+$ is
defined, in the outgoing representation, in terms of support
properties (this is also true for the incoming subspace in the
incoming representation).  One can then easily understand that the 
fundamental difference between Lax-Phillips theory and the standard
quantum theory lies in this property.  The subspace defining the
unstable system in the standard theory is usually defined as the
eigenstate of an unperturbed Hamiltonian, and is not associated with
an interval on a line.  The subspaces of the Lax-Phillips theory are
associated with intervals ({\it i.e.}, the positive and negative
half-lines in the outgoing and incoming representations). It was shown
in ref. 11, for a Lax-Phillips theory constructed on a model of free evolution,
that the generator of the motion restricted to the subspace ${\cal K}$
 is symmetric but not self-adjoint.  To briefly review this point in a
slightly different way, consider the infinitesimal generator of the
semigroup. According to the Stone theorem, the unitary evolution
$U(\tau)$ can be represented as $e^{-iK\tau}$; since $P_+ U(\tau)P_-$
takes ${\cal K}$ into itself, as shown by Lax and Phillips$^{}$, $P_+
U(\tau)P_-= P_{\cal K} U(\tau) P_{\cal K}$, and therefore the
infinitesimal generator (the derivative at $\tau = 0$) satisfies
$$ P_+KP_-= P_{\cal K}K P_{\cal K}.$$
Consider vector-valued functions $f(s),g(s)$ in ${\cal K}$, in the translation 
representation,  so that (we may suppress $P_-$ acting on $f$ or $g$)
$$(P_+K)f(s) =-i\theta(-s) {\partial\over\partial s}f(s).\eqno(3.9)$$
We then take the scalar product with $g$ to obtain
$$ \eqalign{\int_{-\infty}^\infty g^*(s)(P_+Kf)(s)ds &= -i\int
g^*(s)\theta(-s) {\partial f(s)\over \partial s}ds \cr
&= -ig^*(0)f(0) + \int \bigl(-i\theta(-s) {\partial g(s)\over \partial
s}\bigr)^* f(s) ds \cr &= -ig(0)^* f(0) + \int (P_+Kg)^*(s)f(s)ds,\cr}$$
so that the operator $ P_+KP_-= P_{\cal K}K P_{\cal K}$ is not self-adjoint.
 It is through this mechanism, that the subspaces are defined by
(semibounded) support properties on a variable acted upon by
an operator represented as a derivative in the translation
representation (somewhat analogous to the non-self-adjointness of the
operator $i\partial/\partial r$ in three dimensional spherical coordinates), 
 that the Lax-Phillips theory
provides a description that has the semigroup property for the
evolution of an unstable system.
\par It follows immediately from the property of a contractive
semigroup that the generator has, in fact, a family
of complex eigenvalues in the lower half-plane; the eigenfunctions are$^{9}$
$$f_\mu(s) = \cases{e^{\mu s}n, & $s\le 0$;\cr
0, & $s > 0$, \cr} \eqno(3.10)$$
where $n$ is some vector in the auxiliary space.
\par In order to construct an imbedding of Lax-Phillips
theory into the quantum theory$^{13}$
 we assume that there exist {\it wave operators} $\Omega_\pm$ which
intertwine this dynamical operator with an unperturbed dynamical
operator $K_0$.
 We shall assume that $K_0$ has absolutely
continuous spectrum in $(-\infty,\, \infty)$.
\par We begin the development of the quantum Lax-Phillips theory with
the construction of the incoming and outgoing translation
 representations. In this way, we shall
construct explicitly the foliations required by the Lax-Phillips
theory described above$^{13}$.  Let us define the free {\it spectral}
 representation$^{9}$ in terms of the Fourier transform
$${ _f\langle} \sigma \beta \vert g \rangle =  \int e^{-i\sigma s} 
\, {_f\langle} 
 s \beta \vert g \rangle ds , \eqno(3.11)$$
where  $_f\langle \sigma \beta
\vert g \rangle$ satisfies
$$   _f\langle \sigma \beta \vert K_0 \vert g \rangle = \sigma\,_f\langle
\sigma \beta \vert g \rangle . \eqno(3.12)$$
Here $\vert g \rangle$ is an element of $\overline{\cal H}$ and
$\beta$ corresponds to the variables (measure space) of the auxiliary
space associated
to each value of $\sigma$, which, with $\sigma$, comprise a complete
spectral set.  The functions  may be thought of as a set of functions of the
variables $\beta$ indexed on the variable
 $\sigma$ in a continuous sequence of auxiliary  Hilbert spaces
isomorphic to $H$ .  Clearly, $K_0$  acts as
the generator of translations in the representation $_f\langle
 s \beta \vert g \rangle$. 
\par We now proceed to define the incoming and outgoing subspaces
$\cal D_\pm$. 
\par  Let us consider
the sets of functions with support in $L^2(0, \infty)$ and in
$L^2(-\infty,0)$,and call these subspaces $ D_0^\pm$.  The
Fourier transform back to the free spectral representation provides
the two sets of Hardy class functions
$$ _f\langle \sigma \beta \vert g_0^\pm \rangle =  \int e^{-i\sigma
s}\
 _f\langle s \beta \vert g_0^\pm \rangle ds  \in H_\pm , \eqno(3.13)$$
for $g_0^\pm \in D_0^\pm$.
\par We may now define the subspaces ${\cal D}_\pm$ in the Hilbert space of
states ${\overline{\cal H}}$.  To do this we first map these Hardy
class functions in ${\overline H}$ to ${\overline{\cal H}}$, i.e., we
define the subspaces ${\cal D}_0^{\pm}$ by
$$\int \sum_\beta \vert\sigma \beta \rangle_f \  _f\langle \sigma
\beta \vert g_0^\pm \rangle d\sigma \in {\cal D}_0^\pm.  \eqno(3.14)$$
\par   We shall assume that there are wave operators which intertwine $K_0$
with the full evolution $K$, i.e., that the limits
$$ \lim_{\tau \to \pm \infty} e^{iK\tau} e^{-iK_0 \tau} = \,
\Omega_\pm \eqno(3.15)$$
exist on a dense set in ${\overline{\cal H}}$. The conditions for the
existence of such wave operators are satisifed if the difference
between $K$ and $K_0$ is a ``small'' operator.  We shall explicitly
construct these wave operators in the case of the Stark effect that we
study here. 
\par  The construction of ${\cal D}_\pm $ is then completed with the help
of the wave operators.  We define these subspaces by
$$\eqalign{ {\cal D}_+ &= \Omega_+ {\cal D}_0^+  \cr
{\cal D}_- &= \Omega_- {\cal D}_0^-  .\cr} \eqno(3.16)$$
We remark that these subspaces are not produced by the same unitary
map. This procedure is necessary to realize the Lax-Phillips structure
non-trivially; if a single unitary map were used, then there would
exist a transformation into the space of functions on $L^2(-\infty,
\infty, H)$ which has the property that all functions with support on
the positive half-line represent elements of ${\cal D}_+$, and all
functions with support on the negative half-line represent elements of
${\cal D}_-$ in the same representation; the resulting Lax-Phillips
$S$-matrix would then be trivial.
 The requirement that ${\cal D}_+$
and ${\cal D}_-$ be orthogonal is not an immediate consequence of our
construction. Since the functions $_f\langle s \beta \vert g_0^\pm
\rangle$ have support in, respectively, the positive and negative half
lines, and the orthogonality of ${\cal D}_\pm$ is determined by the
integral of the product of these functions with an operator valued kernel
${\bf S}(s-s')$ (to be defined below), one sees that suitable analyticity properties of the
transformed kernel $S(\sigma)$  assure that these subspaces will be 
orthogonal.  This analyticity property (upper half plane analyticity) is true
 in the Stark model that we treat. 
\par The wave operators defined by $(3.15)$ intertwine $K$ and $K_0$, i.e.,
$$ K \Omega_\pm  = \Omega_\pm K_0;        \eqno(3.17)$$
we may therefore construct the outgoing (incoming) spectral representations
from the free spectral representation.  Since
$$ \eqalign{K\Omega_\pm \vert \sigma \beta \rangle_f &= \Omega_\pm K_0 \vert
\sigma \beta \rangle_f \cr
&=\sigma \Omega _\pm \vert \sigma \beta \rangle_f,\cr} \eqno(3.18)$$
we may identify
$$ \vert \sigma \beta \rangle_{out \atop in} = \Omega_\pm \vert \sigma
\beta \rangle_f . \eqno(3.19)$$
The Lax-Phillips $S$-matrix is defined as the operator, on ${\overline H}$,
which carries the incoming to outgoing translation representations of the
evolution operator $K$. Suppose $g$ is an element of ${\overline {\cal H}}$;
its incoming spectral representation, according to $(3.19)$, is
$$ {_{in}\langle} \sigma \beta \vert g) = {_f\langle}\sigma \beta \vert
\Omega_-^{-1} g).  \eqno(3.20)$$
Let us now act on this function with the Lax-Phillips $S$-matrix in the
free spectral representation, and require the result to be the {\it outgoing}
representer of $g$:
$$ \eqalign{{_{out}\langle} \sigma \beta \vert g)
&= {_f\langle} \sigma \beta \vert \Omega_+^{-1} g) \cr
&=\, \int d\sigma'\, \sum_{\beta'}{_f\langle} \sigma \beta \vert
 {\bf S}\vert \sigma'\beta' \rangle_f \,\,
{_f\langle}\sigma' \beta' \vert \Omega_-^{-1} g) \cr} \eqno(3.21)$$
where ${\bf S}$ is the Lax-Phillips $S$-operator
 (defined on ${\overline{\cal H}}$).
Transforming the kernel to the free translation representation
with the help of
$(3.11)$, i.e.,
$$ {_f\langle} s \beta\vert {\bf S} \vert s' \beta' \rangle_f =
{1 \over (2\pi)^2}
\int d\sigma d\sigma' \, e^{i\sigma s} e^{-i\sigma's'}
{_f\langle} \sigma \beta \vert
 {\bf S}\vert \sigma'\beta' \rangle_f , \eqno(3.22)$$
we see that the relation  $(3.21)$ becomes, after using Fourier
transform in a similar way to
transform the {\it in} and {\it out } spectral representations to
the corresponding {\it in} and {\it out} translation representations,
$$\eqalign{ {_{out}\langle} s\beta \vert g) = {_f\langle} s\beta
 \vert \Omega_+^{-1} g) &=
\int ds'\, \sum_{\beta'} {_f\langle} s \beta\vert {\bf S} \vert s'
\beta'
 \rangle_f
\,{_f\langle} s' \beta' \vert \Omega_-^{-1} g) \cr
&=  \int ds'\, \sum_{\beta'} {_f\langle} s \beta\vert {\bf S} \vert s' \beta'
 \rangle_f {_{in}\langle} s'\beta' \vert g). \cr}  \eqno(3.23)$$
Hence the Lax-Phillips $S$-matrix is given by
$$ S= \{ {_f\langle} s \beta\vert {\bf S} \vert s' \beta' \rangle_f
\},
 \eqno(3.24)$$
 in free translation representation. It follows from the intertwining
property $(3.17)$ that
$$ {_f\langle} \sigma \beta \vert{\bf S}\vert \sigma' \beta' \rangle_f =
\delta(\sigma - \sigma') S^{\beta \beta'}(\sigma), \eqno(3.25)$$
 \par This result can be expressed in terms of operators on ${\overline{\cal
H}}$.
  Let
$$ w_-^{-1} = \{ {_f\langle} s\beta \vert \Omega_-^{-1} \}   \eqno(3.26)$$
be a map from ${\overline{\cal H}}$ to ${\overline H}$ in the incoming
 translation
representation, and, similarly,
$$ w_+^{-1} = \{ {_f\langle} s\beta \vert \Omega_+^{-1} \} \eqno(3.27)$$
a map from ${\overline{\cal H}}$ to ${\overline H}$ in the outgoing translation
representation. It then follows from $(3.23)$ that
$$ S= w_+^{-1} w_- , \eqno(3.28)$$
as a kernel on the free translation representation.
 This kernel is
understood to operate on the representer of a vector $g$ in the incoming
representation and map it to the representer in the outgoing
representation.
\bigskip
\noindent
{\bf 4. Application to the Stark model}
\smallskip
\par Since the spectrum of the Stark model that we are using has
spectrum $(-\infty, \infty)$, we may take for the generator of motion
$$ K = H_{Stark}; \qquad K_0 = H_{0\,Stark}; \qquad  V=V_{Stark},
\eqno(4.1)$$
where  $ H_{Stark}, H_{0\,Stark}$ and  $ V=V_{Stark}$ are the operators
defined in Section 2. Since $H_0$ is proportional to $x$, we may make
use of the canonical commutation relations of the quantum theory to
identify the momentum $p$ as proportional to the (foliation) variable of the
unperturbed (free) translation representation, i.e., one may take $s= p/E$,
implying that $[s,K_0]= i$ (we have taken $\hbar =1$).  We then have

$$   (e^{-iK_0\tau}f)(s) = f(s-\tau), \eqno(4.2)$$      
or, in differential form,
$$ (K_0f)(s) = -i{d \over ds}f(s). \eqno(4.3)$$
The auxiliary Hilbert spaces of the corresponding Lax-Phillips theory
are one-dimensional.
\par The spectrum of $K_0$, given by $\{-Ex\}$, with $-\infty < x < \infty$,
can then be identified
with $\sigma$ of the Lax-Phillips (unperturbed) energy
representation. We shall follow this formal identification to develop
the Lax-Phillips theory of resonances, and return to the original
 interpretation of $x$ and $p$ to obtain physical information about
 the resonant state.
\par The wave operators are defined as 
$$ \Omega_\pm = \lim_{\tau \to \pm \infty} e^{iK\tau} e^{-iK_0\tau}. 
 \eqno(4.4)$$
\par We shall calculate the matrix elements of the wave operator in
the unperturbed energy representation. It will be convenient,
moreover, to use directly the measure on the spectrum of $x$; we
therefore use kets of the form $\vert x\rangle \equiv \sqrt{|E|}\vert
\sigma\rangle $.
\par Following the standard procedure for taking these limits ,
one finds for the representation $\{\vert x\rangle\}$, that
$$ \langle x \vert \Omega_\pm \vert x' \rangle = \delta (x-x') -
\lim_{\varepsilon \to 0_+}
\langle x \vert {1 \over H+ Ex' \pm i\varepsilon}V \vert
x'\rangle. \eqno(4.5)$$ 
Since this formula is bilinear in the kets $\vert x \rangle$ and
$\vert x'\rangle$, one could use equally well the kets $\vert \sigma
\rangle$ and $\vert \sigma' \rangle$.
\par The operator multiplying $V$ in $(4.5)$ is $-G(z)$, for
$z=-Ex'\mp i\varepsilon$; the matrix elements of this operator were
evaluated in $(2.8)$.  Carrying out the integral for the product
$G(z)V$ with the help of the definition $(2.7)$, one finds that
$$ \langle x \vert  G(z)V\vert x' \rangle =  \lambda \sqrt{2 \over
\pi} {e^{-(x^2 + x'^2)} \over (z+Ex)( 1- \lambda \sqrt{2 \over \pi}
F(z))} \eqno(4.6)$$
The wave operators are then given by
$$ \langle x \vert \Omega_\pm \vert x' \rangle = \delta(x-x') +
 \lambda \sqrt{2 \over
\pi} {e^{-(x^2 + x'^2)} \over (E(x-x')\mp i\varepsilon)( 1- \lambda
 \sqrt{2 \over \pi}F(-Ex' \mp i\varepsilon))}\eqno(4.7)$$
 Using a partial fraction
decomposition for the product of the $GV$ terms, one easily verifies that
the operators $\Omega_\pm$ are unitary.
\par We now turn to the construction of the incoming and outgoing
translation representations. To do this, we follow the method of reference 13. 
We define the free outgoing translation as the set of functions with
support in $s$ (i.e., $p/E$ in the Stark model) on the positive real
axis.  By $(3.11)$, the functions
$$ f_+^0 (x) = \int_0^\infty e^{ipx}f_+^0(p) dp \equiv \langle x \vert
f_+^0\rangle \eqno(4.8)$$
are in the free outgoing translation representation, and are analytic
in the upper half $x$-plane (lower half $\sigma$-plane).  Since the
wave operators intertwine $K_0$ and $K$, functions of
the (full) outgoing representation are then given by
$$ \int_{-\infty}^\infty \, \langle x' 
\vert \Omega_+ \vert x \rangle dx f_+^0(x) dx = f_+^{out}(x') \in
D_+. \eqno(4.9)$$
\par Given a function of the type $f_+^0(x)$, one can calculate
the resulting function $f_+^{out}(x')$ explicitly by noting that the
boundary value of $F(z)$ from below the real axis is 
$$ F(-Ex')_{below} = \lim_{\varepsilon \to 0_+}\int_{-\infty}^\infty 
{e^{-2x''^2} \over
E(x''-x')-i\varepsilon}dx'' = {i\pi \over E}
e^{-2x'^2} erfc\bigl(-i\sqrt{2}x' \bigr) \eqno(4.10)$$
\par For the construction of the incoming translation representation,
one uses $\Omega_-$ and a corresponding set of functions $f_-^0(x)$
with support on the negative half line.  The kernel of integration
then contains $ F(-Ex' - i\varepsilon)$; this may be obtained from
$(2.10)$, i.e., 
$$ F(-Ex')_{above}= F(-Ex')_{below} - {2\pi i \over E}
e^{-2x'^2}. \eqno(4.11)$$ 
\par We now turn to the calculation of the $S$-matrix.  We see from
$(3.21)$ that 
$$ \langle x \vert {\bf S} \vert x'\rangle =  \langle x
 \vert \Omega+^{-1} \Omega_- \vert x'\rangle. \eqno(4.12)$$
We now use the definition $(3.15)$ for the wave operators; following
the standard method$^{22}$, we find that 

$$\langle x \vert {\bf S} \vert x'\rangle = \delta (x-x')( 1 + {2\pi i
\over E} \lim_{\varepsilon \to 0_+}\langle x \vert T(-Ex' +
i\varepsilon)\vert  x'\rangle, \eqno(4.13)$$
where  $T(z) \equiv V(1+ G(z)V)$. We now compute 
$$ \eqalign{\lim_{\varepsilon \to 0_+} \langle x \vert
T(-Ex'+i\varepsilon)\vert x' \rangle &= \lim_{\varepsilon \to 0_+}
\int_{-\infty}^\infty  \lambda
\sqrt{2\over \pi} e^{-x^2 + x''^2} \bigl[ \delta(x'' - x')\cr &-
{ \lambda \sqrt{2 \over \pi} e^{-(x'^2 + x''^2)} \over (E(x'-x'')
-i\varepsilon)( 1 - \lambda\sqrt{2\over \pi}F(-Ex' +i\varepsilon)}\bigr]dx''\cr
&= \lim_{\varepsilon \to 0_+}
{ \lambda\sqrt{2\over \pi}e^{-(x^2 + x'^2)}\over 
 1 - \lambda\sqrt{2\over \pi}F(-Ex' +i\varepsilon)}.\cr} \eqno(4.14)$$
  \par It then follows that 

$$\langle x \vert {\bf S} \vert x'\rangle= \delta(x-x')\bigl[ 1+ {2\pi
i \over E}\lim_{\varepsilon \to 0_+}
{ \lambda\sqrt{2\over \pi}e^{-2x^2}\over 
 1 - \lambda\sqrt{2\over \pi}F(-Ex +i\varepsilon)}\bigr], \eqno(4.15)$$
so that we may write
$$\langle x \vert {\bf S} \vert x'\rangle \equiv \delta(x-x')S(x).
\eqno(4.16)$$     
\par If we write the semigroup evolution, restricted to the subspace
${\cal K}$, as 
$$ Z(\tau) = e^{-iB\tau}, \eqno(4.17)$$
it follows from the contractive semigroup property that the operator
$B$ has an eigenvector in the outgoing representation satisfying
$$ Bf_{out} = \mu f_{out}, \eqno(4.18)$$
 with $\mu$ in the lower half plane, for which the eigenfunctions are
of the form (with support in $(0, \infty)$)
$$ f_{out}(s) = e^{-i\mu s} n, \eqno(4.19)$$
where $n$ is a vector in the auxiliary space (in our Stark model, one
dimensional).  The eigenfunctions in the outgoing $x$ representation 
(corresponding to the ``energy'' variable $\sigma$), are then of the form
$$ f_{out}(x)  ={in \over x-\mu}. \eqno(4.20)$$
The $S$-matrix, connecting the incoming to outgoing representations,
therefore has the form
  $$ S(x) \sim {r \over x-z_0}, \eqno(4.21)$$
where $z_0$ is the position of the pole of the diagonal $S$-matrix in
the lower half plane, which we identify with the semigroup exponent $\mu$,
and $r$ is the residue.  From $(4.15)$, one sees
that the pole of the $S$-matrix corresponds to a zero of the denominator 
$$1 - \lambda\sqrt{2\over \pi}F(-Ex +i\varepsilon)$$
continued to the lower half plane, i.e., we must find the zero of 
$$1 - \lambda\sqrt{2\over \pi}F^\ell(z). \eqno(4.22)$$
This is precisely the pole (zero of $(2.15)$) which occurs in the pole
 approximation of the Wigner-Weisskopf theory discussed in Section 2.  
The residue of the pole in $(4.21)$ is then given by 
$$ r = {8\pi i\over E^3}\lambda \sqrt{2 \over \pi}{e^{{-2z_0^2\over
E^2}}\over z_0 -\lambda}. \eqno(4.23)$$

\bigskip  
 \noindent
{\bf 5. Conclusions and Discussion}
\smallskip
\par We have studied a model for resonances which provides the
 possibility of analysis using the Wigner-Weisskopf method, the
standard technique for such analyses, giving exponential behavior in
the pole approximation, but also accessible to direct analysis in the
Lax-Phillips framework.  In the Wigner-Weisskopf analysis, the
resonance is described by the position of a pole in the complex
energy plane, and does not have a state in the Hilbert space
associated with it.  Although there has been considerable study of
the application of the method of rigged Hilbert spaces$^{23}$, or Gel'fand
triples ($D \subset H \subset {\overline H}$, where ${\overline H}$ is
defined as the set of bounded linear functionals on the subspace $D$),
 for the description of resonances which satisfies the
property of exact exponential decay (without the ``background''
corrections to the Wigner-Weisskopf pole approximation), the elements
of the Gel'fand triple are not, in general, vectors of a Hilbert
space (they belong to a Banach space), and have no scalar products.
Hence, it is not possible to compute the expectation value of an
observable (unless its extension to $\overline H$ maps it into the
smaller space $D$ on which the linear functionals of $\overline H$ are
defined), or to study physical properties such as localization of the state. 
\par  The
Lax-Phillips formulation describes the resonant state as an element of
a Hilbert space, and answers to such questions then become
accessible. As a simple example, we have computed the
 resonant state for the
Stark model that we have used in $(4.20)$.  The variable $x$ used
here corresponds to the ``energy'' in the Lax-Phillips formal
structure, but retains its physical meaning in the result as position.
 Note that this is also true in our formulation of the
Wigner-Weisskopf model, taking for the unperturbed Hamiltonian the
term $-Ex$ in the Hamiltonian $(2.1)$ since this term is large
compared to the term producing the imbedded bound state.  The position
 variable occurs in the Hamiltonian producing a continuous energy
spectrum $\{Ex\}$.  The interpretation of the poles
of the $S$-matrix, or resolvent therefore remains, as in the usual
formulation of resonance problems, as occurring in the complex energy
plane, but the variable $x$ retains its physical meaning as coordinate
as well.
 This method leads us to an approximate exponential decay law
in time in the pole approximation of Wigner-Weisskopf theory, and to
an exact exponential decay law in the Lax-Phillips treatment, with
precisely the same exponent. 
\par  In the
framework of general Lax-Phillips theory, the resonant state carries
the pole as in $(4.20)$ where the $x$ that appears there would be
replaced by another symbol, say, $\sigma$, associated with unperturbed
energy, and
the distribution over space of the wave function would reside in the
vector $n$ of the auxiliary Hilbert space.  In our case, this vector
is just a number (one-dimensional), and the space distribution is
provided by the equivalence of (unperturbed) energy and the variable $x$.
\par The resonance state provided by the Lax-Phillips theory in the
``energy'' representation actually therefore corresponds to a distribution 
 of $x$ values in the resonant state.
$$ \vert f_{out}\vert^2 = {\Vert  n \Vert^2 \over (x- Re\, z_0)^2 +
\vert Im\,z_0\vert^2}, $$
a Cauchy distribution with width $\vert Im\,z_0 \vert$.  The Cauchy
distribution is centered on $Re z_0$, corresponding to a shift away
from the mean value of $x=0$ in the bound state in the absence of
electric field; as we have seen, the pole moves further to the left
with increasing field, so that the center of the wave packet moves to
the left.

\bigskip
\noindent
{\bf References}
\frenchspacing
\item{1.} V.F. Weisskopf and E.P. Wigner, Zeits. f. Phys. {\bf 63},
(1930); {\bf 65}, 18 (1930).
\item{2.} B. Misra and E.C.G. Sudarshan, Jour. Math. Phys. {\bf 18},
756 (1977). For more recent discussion and references, see P. Facchi
and S. Pascazio, ``Unstable Systems and Quantum Zeno Phenomena in
Quantum Field Theory,'' quant-ph/0202127.
\item{3.} N. Bleistein, R. Handelsman, L.P. Horwitz and H. Neumann,
Nuovo Cimento {\bf 41A}, 389 (1977).
\item{4.} L.P. Horwitz and J.-P. Marchand, Helv. Phys. Acta {\bf 42},
1039 (1969); L.P. Horwitz and J.-P. Marchand, Rocky Mountain Journa;l
of Mathematics {\bf 1}, 225 (1971).
\item{5.} C. Piron, {\it Foundations of Quantum Physics},
W.A. Benjamin, Inc., Reading, Mass. (1967); see also discussions in
ref. 11.
\item {6.} B. Winstein, {\it et al, Results from the Neutral Kaon
Program at Fermilab's Meson Center Beamline 1985-1997,\/} published on
behalf of the E731, E773 and E799 Collaborations, Fermi National
Laboratory, P.O. Box 500, Batavia, Illinois.
\item{7.} T.D. Lee, R. Oehme and C.N. Yang, Phys. Rev. {\bf 106}, 340
(1957); T.T. Wu and C.N. Yang, Phys. Rev, Lett. {\bf 13}, 380 (1964).
\item{8.} L.P. Horwitz and L. Mizrachi, Nuovo Cimento {\bf 21A},
625(1974);  E. Cohen and L.P. Horwitz, Hadronic Jour. {\bf 24} 593
(2001); see also hep-th/9808030, hep-ph/9811332.
\item{9.} P.D. Lax and R.S. Phillips,{\it Scattering Theory}, Academic
Press, N.Y. (1967).
\item{10.} C. Flesia and C. Piron, Helv. Phys. Acta {\bf 57}, 697
(1984).
\item{11.} L.P. Horwitz and C. Piron, Helv. Phys. Acta {\bf 66}, 694
(1993).
\item{12.} E. Eisenberg and L.P. Horwitz, in {\it Advances in Chemical
Physics\/} vol.XCIX, p. 245, ed. I. Prigogine and S. Rice, Wiley, New York
(1997).
\item{13.} L.P. Horwitz and Y.Strauss, in {\it Irreversibility and
Causality, Semigroups and Rigged Hilbert Spaces \/}, ed. A. B\"ohm,
H.-D. Doebner and P. Kielanowski, Springer Verlag, New York (1996). 
\item{14.} Y. Strauss, Int. Jour. Theor. Phys. {\bf 42}, 2285 (2003).
\item{15.}   W. Baumgartel, Math. Nachr. {\bf 69}, 107 (1975);
 L.P. Horwitz and I.M. Sigal, Helv. Phys. Acta
 {\bf 51}, 685 (1978); G. Parravicini, V. Gorini
 and E.C.G. Sudarshan, J. Math. Phys. {\bf 21}, 2208
 (1980); A. Bohm, {\it Quantum Mechanics: Foundations
 and Applications\/,} Springer, Berlin (1986); A. Bohm,  M. Gadella
and
 G.B. Mainland, Am. J. Phys. {\bf 57}, 1105 (1989); T. Bailey and
 W.C. Schieve, Nuovo Cimento {\bf 47A}, 231 (1978).
\item{16.} O. Agam, N.S. Wingreen, B.L. Altshuler, D.C. Ralph, and
M. Tinkham, cond-mat/9611115; O. Agam, B.L. Altshuler and
A.V. Andreev, cond-mat/9509102; B. Altshuler, {\it From Anderson
Localization to Quantum Chaos},
Lecture at the Institute for Advanced Study, Princeton, N.J.,
 7 February, 1996. 
\item{17.} E.C.G. Stueckelberg, Helv. Phys. Acta {\bf 14}, 322, 588
(1941); J. Schwinger, Phys. Rev. {\bf 82}, 664 (1951); R.P. Feynman,
        Rev. Mod. Phys. {\bf 20}, 367 (1948) and Phys. Rev. {\bf 80},
        440 (1950); L.P. Horwitz and C. Piron, Helv. Phys. Acta {\bf
        46}, 316 (1973); R. Fanchi, Phys. Rev {\bf D20},3108 (1979);
        A. Kyprianides, Phys. Rep. {\bf 155}, 1 (1986) (and references
 therein).
\item{18.} Y. Strauss and L.P. Horwitz, Found. Phys. {\bf 30}, 653 (2000).
\item{19.} K.O. Friedrichs and P.A. Rejto, Comm. Pure and
Appl. Math. {\bf 15}, 219 (1962).
\item{20.} I.P. Cornfield, S.V. Formin and Ya. G. Sinai,
{\it Ergodic Theory}, Springer, Berlin (1982).
\item{21.} Y. Strauss, E. Eisenberg and L.P. Horwitz,
Jour. Math. Phys. {\bf 41}, 8050 (2000).
\item{22.}  For example, J.R. Taylor, {\it Scattering Theory},
 John Wiley and Sons,
 N.Y. (1972); R.J. Newton, {\it Scattering Theory of Particles
 and Waves}, McGraw Hill, N.Y. (1976).

\vfill
\end
\bye